# Measurements of the Atmospheric Electric Field through a Triangular Array and the Long-range Saharan Dust Electrification in Southern Portugal

H.G. Silva,[1*] F. Lopes,[1] S. Pereira,[2] S.M. Barbosa,[3] K. Nicoll,[4] M. Collares Pereira,[1] R.G. Harrison,[4]

[1]*Renewable Energies Chair, University of Évora, IIFA, Palácio do Vimioso, Largo Marquês de Marialva, 7002-554, Évora, Portugal. (*hgsilva@uevora.pt)*

[2]*Departamento de Física, ICT, Instituto de Ciências da Terra, Universidade de Évora, Rua Romão Ramalho 59, 7002-554 Évora, Portugal.*

[3]*INESC TEC - INESC Technology and Science, Campus da FEUP, Rua Dr Roberto Frias, 4200-465 Porto, Portugal.*

[4]*Department of Meteorology, University of Reading, Berkshire, RG6 6BB, UK.*

**Abstract**

Atmospheric electric field (AEF) measurements were carried out in three different sites forming a triangular array in Southern Portugal. The campaign was performed during the summer characterized by Saharan dust outbreaks; the 16[th]-17[th] July 2014 desert dust event is considered here. Evidence of long-range dust electrification is attributed to the air-Earth electrical current creating a positive space-charge inside of the dust layer. An increase of ~23 V/m is observed in AEF on the day of the dust event corresponding to space-charges of ~20-2 pCm$^{-3}$ (charge layer thicknesses ~10-100 m). A reduction of AEF is observed after the dust event.

# 1. Introduction

Dust storms have been receiving significant attention in the past decades (e.g., Engelstaedter et al., 2006), among different roles, because of their impact on the planetary radiactive forcing and its relevance in Earth's climate. Though little information has been collected on dust electrification (e.g. Ette, 1970, Ulanowski et al., 2007), the interest has been raised recently due to its importance in two main areas: energy systems and planetary exploration. In the former, dust electrification can have technological importance since it is of great usefulness in the development of automatic electrostatic dust particle removal from Solar Energy systems, as it was used on lunar missions (Calle et al., 2009). This technological improvement on Earth will permit the increase of the efficiency in these systems while reducing water consumption (Sarver et al., 2013). In the latter, the understanding of Martian dust devils electrification (Delory et al., 2006) is expected to be boosted by the ExoMars mission (Esposito et al., 2014) which will deploy later this year two payloads: DREAMS and MicroARES; these measuring instruments are expected to further contribute to the understanding of these phenomena on Mars. Moreover, Williams et al. (2009) reported on the electrification of haboobs in Sahelian belt of West Africa. Measurements were made in the (source) region where the storms developed and significant electric perturbation where only found under heavy dust (high concentration of large sized particles) exhibiting in most of the occurrences strong and negative monopolar electrifications (absolute electric fields of ~1-10 kV/m). These observations seem to be line with the ones by Kamra (1972); in which the author states that most of the dust storms dominated by clay minerals tend to produce negative space charges in the source region. Nevertheless, debate still exists on ether the negative electrification comes from clay minerals (dust) or quartz minerals (sand), Williams et al. (2009).

Many open questions also exist on the way space charge generated in the source region behaves under long-range transport. In principle, only small particles (e.g., clay particles with size ranges from 1 to 100 μm) can be transported and according to previous observations of negative clay electrification, negative perturbations in the electric fields would have to be seem away from the source region. Even so, the detailed work of Reiter and co-authors (Reiter, 1992) seems to contradict this. The author has shown that, during a Saharan dust outbreak that reached the Zugspitze Peak (Germany), a positive space charge density (SCD) at ~3 km altitude was formed two times higher than in the normal "clean" conditions. Tropospheric LIDAR measurements showed that the dust layer was co-located with the space charge density around ~3 km and chemical aerosol analysis showed that sand particles were dominant with significant increased concentrations of $SiO_2$ and $Al_2O_3$. More recently, balloon-borne charge measurements of Saharan dust layers (up to 4 km) have been made in Cape Verde Islands, where Saharan dust outbreaks frequently occur. The experiment depicted a maximum positive charge density of ~ 25 pC m$^{-3}$ (Nicoll et al., 2011). Furthermore, the authors estimated that dust charge takes roughly 70 s to decay (Nicoll et al., 2011) and consequently no long-range electrification would be observable. For that reason, the authors argued that a possible mechanism to explain long-range dust charging was the vertical air-Earth electric current, imposed by the global electrical circuit (GEC), flowing through the atmospheric electric conductivity gradient inside the dust layer (Nicoll and Harrison, 2010). The atmospheric electric conductivity gradient is a consequence of small ion scavenging by dust particles. In which the ion-particle attachment process charges dust particles (with low electrical mobilities) that significantly decrease electric conductivity (Ulanowski et al., 2007) and generates the conductivity gradient. Ohm's law will then reflect this reduction in the conductivity by an increase in the Atmospheric Electric Potential Gradient (PG[1]). Thus, if dust charging was caused by the GEC action, it would imply that dust desert plumes would be positively charged far away from its source and positive perturbations on the PG should be found. This was the case of Reiter (1992) observations and the present ones also tend to corroborate this.

Previous work on the long-range dust electrification has been focused on a single measuring site where the PG was recorded (e.g., Rudge, 1913). Nevertheless, recent efforts in atmospheric electricity concern the development of networks of PG field-mills in large time (~1 hour) and space scales (~100 km), as it is the case of the network installed in South America (Tazca et al., 2014) and the one under development in Europe. The existence of such networks raises the possibility of the use of coordinated PG measurements to track atmospheric phenomena such as smoke plume transport, known to affect PG measurements (Conceição et al., 2015). In this context, an experiment was conceived and undertaken during the ALEX2014 meteorological campaign (www.alex2014.cge.uevora.pt). It consisted on the installation of three similar PG field-mills, in Southern Portugal, forming a triangular array that allowed the recording of PG time series during a three-month period, from June to August 2014. This period corresponds to the summer season in the northern hemisphere and represents a unique opportunity to perform such experiment due to two main reasons: the frequency of occurrences of fair-weather days and the occurrences of isolated Saharan dust outbreaks transported over Africa to the measuring region (Obregón et al., 2015). The use an array of sensors instead of a single sensor is because it should permit to distinguish global perturbations from local ones.

---

[1] In Atmospheric Electricity it is common to use PG, as means to quantify the Atmospheric Electric Field. The convention is that the PG is defined by $PG = dV_I/dz$, where $V_I$ is the ionospheric potential with respect to Earth's surface (where $V=0$) and $z$ is the vertical coordinate. By this convention the PG is positive for fair-weather days (according to the international standards fair-weather days are selected as those with cloudiness less than 0.2, wind speed less than 5 ms$^{-1}$ and with the absence either of fog or precipitation, Chalmers, 1967) and related to the vertical component of the atmospheric electric field $E_z$ by $E_z = -PG$. GEC is a consequence of the $V_I$ (Rycroft et al., 2000), it is charged in the thunderstorm active regions of the globe and discharged in the fair-weather regions by the flow of an air-Earth electric current (Conceição and Silva, 2015). The daily variation of thunderstorm activity modulates globally the PG in what it is called the Carnegie Curve (Harrison, 2013).

This paper is organized as follows: section 2 describes the experimental setup, section 3 outlines the Saharan dust event of July 16$^{th}$ 2014 (day 46 of the campaign); section 4 presents the PG measurements during the ALEX2014 campaign; section 5 discusses the results and a brief formulation is derived to reinforce the observations; and in section 6 main conclusions along with recommendations for future work are given.

## 2. Potential Gradient (PG) array and Aerosol Optical Depths (AOD) measurements

An equilateral triangle is formed by three JCI field-mills (Chubb, 2014; Chubb, 2015), separated by nearly 50 km from each other, forming an triangular array of about ~1000 km$^2$ in Southern Portugal (Figure 1). The geographic location of the three sites in which measurements of PG were conducted are: Évora (EVO) at 38.50 N, 7.91 W; Amieira (AMI) at 38.27 N, 7.53 W and Beja Airbase (BEA) at 38.07 N, 7.93 W. The EVO and BEA sites follow almost a North-South alignment, whilst AMI is more deviated to the East and is settled approximately in the mid-way of the other two sites. The EVO station is situated in the center of the city of Évora (~50 000 inhabitants), where major sources of pollutants are due to anthropogenic activity such as traffic, heating (winter) and cooling (summer) air systems. In EVO, a JCI 131 was installed in the University of Évora campus (at 2 m height) with few trees and two University buildings in its surroundings (~50 m away). The instrument was calibrated in 2012 and has been operating since 2005. The AMI station is located on the shoreline of the Southern part of the Alqueva reservoir (currently the largest man-made lake in Western Europe), set upon a hill approximately 30 m above the lake water level, with low vegetation in its surroundings (Lopes et al., 2015). The BEA station is located further south on an air base in the outskirts of the small city of Beja (~40 000 inhabitants). In AMI and BEA two identical field-mills JCI 131F were used and installed as well at 2 m height above the ground. The characterization of the aerosol conditions in the region was based on the AERONET station (Holben et al., 1998,) located at EVO. An automatic sun tracking photometer (CIMEL CE-318-2) is operated to measure aerosol optical depths (AOD) at several wavelengths in the range of 340-1640 nm. The AOD are a measure of the solar radiation extinction due to the aerosol load present in the atmospheric column. Moreover, the spectral dependence of the optical depth, AE (Angstrom exponent), provides information on the size distribution of the aerosol population (i.e., aerosol fine and coarse modes relative proportion).

## 3. Desert Dust Transported into Southern Portugal

During the period of the study, the presence of Saharan dust over the campaign region was detected by sun-photometer with maximum intensity on 16$^{th}$-18$^{th}$ July. Trajectory analysis (not shown) confirmed this scenario of dust transported from the Sahara region. Various sun-photometer measurements within AERONET network, including the one installed at Évora (EVO), are depicted in Figure 2, which provides some further insight on the dust outbreak, including an indication on its spatial extension. The measurements permit to perceive dust plume extended at least up to central Iberia Peninsula. One of the stations, Badajoz in Spain (BJZ), is located near Évora, while the other three are clustered in the south of Spain, Málaga (MLG), Granada (GRA) and Cerro de Poyos (CDP) (see Figure 1). Optical depths up to 0.51 (at 440 nm) and small wavelength dependence of the optical depth, a typical signature of dust, where observed in the period around 16$^{th}$-18$^{th}$ July, as can be observed in Figure 2. A stronger increase in the aerosol perturbation can be seen in AOD, during 16$^{th}$ July, which persisted during the following two days; on 19$^{th}$ July the perturbed aerosol was no longer visible in the data. Data is scarce for EVO and BJZ on 18$^{th}$ and 19$^{th}$ July, due to clouds. The NAAPS maps in Figure 3 are consistent with this picture. A north-eastward movement apparent in Figure 3 is fairly consistent with the strong increase in optical depth at all sites in the beginning (16$^{th}$ July). Additionally some south-to-north gradient in aerosol load is observed, as the optical depths measured in the south of Spain stations were always higher if compared to Évora and the neighboring Badajoz station. It can be concluded that the sites were the electric field was being measured were under the dust perturbation. It's interesting to notice that on the one hand CDP (next to Granada station) is a mountain site i.e., the measurements are performed near 2000 m (1830 m a.s.l.); on the other hand the optical depths measured here are a large fraction, about 60-75%, of the optical depths measured in GRA (680 m a.s.l.) and MLG (40 m a.s.l.). This means that the dust was mainly present in the free troposphere, which most frequently happens and is known in the literature (e.g., Preißler et al., 2013). Bearing in mind a mean layer height of 3.7 km for Saharan dust layers in the free troposphere over Évora, after Preißler et al. (2013), it is fair to assume that during this episode the dust could also be mainly found in elevated layers in the region where the present study was conducted. Information from the satellite-Borne Lidar CALIOP (Winkler et al., 2007) provides further insight on the dust plume and seems to confirm the above discussion. Its ground track was near the Portuguese coast, at less than 200 km from the region under study and close to 03:00 UTC. In figure 4, the attenuated backscatter coefficient indicates the presence of a layer aloft, between about 2 and 3 km, for the region under study, and its high depolarization capability, as given by the depolarization ratio, confirms the dusty nature of the aerosol plume observed.

## 4. Potential Gradient data

To easy the understanding of the plots it is set the beginning of the campaign as day 1 (1$^{st}$ June 2014) until day 88 (28$^{th}$ August 2014). In this notation the dust event of 17$^{th}$ of July corresponds to day 46 of the campaign. Pollution levels in the BEA and AMI are low in comparison to EVO. This is a fundamental aspect on the present study since high pollution levels leave a common signature on the PG records in large metropolis (Silva et al., 2014). A quality control criterion for the data was used, and values within the precision threshold of field-mills (~|1| V/m) were rejected. This allows the removal of values that correspond to equipment malfunction and-or maintenance, such as when a field-mill stops operating but the data logger continues to record. Two analyses were performed: a robust lowess (locally weighted linear regression) smoothing and a wavelet analysis over the 1-hour averaged data. The wavelet periodogram is, in some sense, a visual description of the way the dominant periods on the data developed during the campaign.

The PG measurements in EVO station are represented in Figure 5a, where a 13-day gap is shown due to equipment malfunction (starting on 12$^{th}$ June at 00:00 UTC until 25$^{th}$ June at 00:00 UTC). An apparent modulation of the PG can be seen with the lowess curve after the desert dust event on day 46 (marked as a dashed vertical line). It seems to point towards a reduction of the PG after the dust event that is observed in the other two stations and which can be the result of a space charge generated by the desert dust event itself (as will be further discussed). This could establish a signature of the event in the PG data, though it should be mentioned that the dip does not start at the same time as the dust appears in the AOD measurements. Here, the vertical line corresponds to the maximum of dust concentration, with the dust event starting two days before; no noticeable change in the PG was observed at the starting day, though. This could be related to the time needed for enough dust to accumulate in the column above in order to affect the PG. Additionally, it is represented in Figure 5b the wavelet periodogram of the PG at EVO, the dominant periods are marked with solid black contour lines. This figure shows the persistence of the one-day periodicity throughout the observational period, as expected. The one-day periodicity is a consequence of the action of the global electrical circuit and the absence of that periodicity could mean that the PG is being perturbed. This is the case for the day of the dust event, day 46.

The PG measurements in AMI were previously discussed on Lopes et al. (2015) in the context of radon interaction with atmospheric ions in fair-weather conditions. In the present analysis the complete record of the time-series at Amieira is used and depicted in Figure 6a. Here, no gaps on the data are observed. Additionally, the PG in AMI shows strong oscillations that occurred on day 22 of the campaign and which were probably caused by thunderstorms and periods of heavy rain. This event has a specific signature on the wavelet analysis (Figure 5b), presenting high spectral power from periods of 2-hours up to 2 days. A similar variation is found in the BEA station. Unfortunately, no data is available in the EVO station on that day. Moreover, the wavelet periodogram for AMI, Figure 6b, also shows the persistence of the one-day periodicity, but besides it evidences a half-day periodicity that could reflect the action of the lake near to which the station was installed, Lopes et al. (2016). A clear one-day periodicity is present in the day of the dust event.

The PG measurements in BEA station (Figure 7a) and the correspondent wavelet analysis (Figure 6b) are generally similar to those observed in AMI. Nevertheless, BEA data shows two sizable gaps: a smaller one from the 1$^{st}$ of June 13:50 UTC until the 3$^{td}$ of June 12:55 UTC and a larger from the 3$^{td}$ of July 22:47 UTC up to the 9$^{th}$ of July 9:35 UTC. Furthermore, the most significant information extractable from the wavelet periodogram for BEA, Figure 7b, is the persistence of the one-day periodicity and that it appears to be diminished during the desert dust event.

In terms of the global effect of the desert dust event (marked by a vertical dashed line) on the PG, it is visible the same modulation of the lowess curve in the EVO station (Figure 4a), with a tendency to a small reduction after the event, in BEA (Figure 5a) and AMI (Figure 6a) stations. This can be a confirmation that the desert dust signatures are not local, instead they result from a regional process. In this spirit, the NAAPS maps (Figure 3b,c) depict that the desert dust event arrived at the three stations simultaneously in the time scale of the dust plume transport (from hours to days), which means that the influence of this event should be similar and closely synchronous in all stations. It should be pointed out that the PG data from BEA station shows apparently an increase in the PG on the day of the event, followed by a decrease of the trend in the days after the desert dust event. This is highlighted by the lowess smoothing (Figures 4a, 5a and 6a).

To deepen into the possible effects of the Saharan dust on the PG measurements done by the array of sensors, it is represented the daily variation for EVO, AMI and BEA stations for days 44 to 48 of the ALEX2014 campaign, respectively, in the upper, middle and lower five panels of Figure 8. The thick black lines in the plots represent the lowess smoothed mean daily variation (i.e., the mean daily cycle with a lowess curve over it) of the PG for the all campaign in the respective stations to be used as comparison. It should be mentioned that the Saharan dust event started on day 45 (15$^{th}$ July 2014) and ended on day 47 (17$^{th}$ July 2014), having full impact on days 46-47. Clear perturbations are seen in the PG plots for day 47 when the PG significantly departs from the mean daily behavior in the three stations. Distortions can be appreciated during the all day at EVO, values reaching ~200 V/m during noon at AMI and strong oscillations in the morning of day 47 at BEA station. In the day after the Saharan dust event, day 48, the PG curves for EVO and AMI remarkably follow the mean behavior and BEA follows fairly the mean curve. Days 44, 55 and 46 of the campaign have similar PG variations in each station and any short-term perturbation of related with the event is hardly noticeable. Exception is made for the day 44 at BEA station where it is visible a big negative dip in the afternoon, clearly different from the oscillations seen on day 47. Finally, days 44 to 47 have undisturbed global radiation curves in EVO and that is considered as reference to the two other locations. The day 48 has disturbed radiation curves consistent with the lack of AOD data. Thus perturbations in the PG curves for the days 44 to 47 cannot be attributed to cloud passage.

## 5. Discussion

In order to quantify the possible impact of the desert dust layer on the PG, the difference between the observed PG during the event and the expected PG without the dust layer (defined as $\Delta F$) is estimated through the following described procedure. The PG data in three stations and the AE are presented in a daily boxplot[2] representation (Figure 8) around the time of occurrence of the maximum of desert dust event (day 46 of the campaign). A significant reduction of the AE is clearly depicted (Figure 8a), identifying the aforementioned event. Both AMI and BEA stations highlight an increase in the median from the lowess trend that is not observed in EVO. The fact that this increase is not present in EVO can be a result of local influences affecting the PG, since the EVO station is located in an urban environment, affected by local pollution (mainly from traffic) which is known to impact severely on the PG (Silva et al., 2014). In fact, the desert dust plume is expected to travel around altitudes of about 3 km from the surface (Obregón et al., 2015), implying that atmospheric processes below it (specially below the planetary boundary layer, PBL) can easily disguise the desert dust effect. Furthermore, if the lowess median value for the day of the event on AMI and BEA is used as the long-term reference for clean days behavior, 57.8 and 76.8 V/m (respectively), the median value for that day on both stations, 80.5 and 100.6 V/m (respectively), can be considered as the value for the perturbed PG. These allow the estimation of the observed $\Delta F$, 22.7 and 23.8 V/m, for AMI and BEA (respectively). The $\Delta F$ value is remarkably similar in both stations and reveals a possible connection in the way that both stations react to the desert dust event as long as they are not affected by local effects. It should be mentioned that using the lowess trend as a reference is a way to quantify the possible effect of the desert dust electrification and is not mentioned to give a precise result. For example, on day 52, Figure 8, a similar situation is found though not attributable to the desert dust event here reported.

---

[2] On each box, the central dot is the median, the limits of the box are the $25^{th}$ (first quartile, $q_1$) and $75^{th}$ (third quartile, $q_3$) percentiles and the whiskers (solid lines) extend to the most extreme data points not considered outliers. Maximum whisker length (w) is set to 1.5 and outliers are defined to be larger than $q_3 + w(q_3 - q_1)$ or smaller than $q_1 - w(q_3 - q_1)$.

Moreover, the detailed view on Figure 8 also highlights the trend for PG decrease after the desert dust event, in particular on day 48 of the campaign (two days after the event). This phenomenon can be attributed to the dispersion of the space charge in the low electric conductive region below the PBL, after the plume has passed. It should be mentioned that the dust event does not start at the same time as the dust appears in the AE measurements, but rather a few days before, however there is no noticeable change in the PG at the time. One explanation should be attributed to the time that is necessary for enough dust to accumulate in the column above in order to affect the PG.

Dust electrification is usually recognized to result from contact and triboelectric charging between particles being bowled. The basic mechanism for charge separation is commonly thought to be the fact that, during collisions, the smallest grains gain negative charge with respect to larger particles (Freier, 1960; Inculet et al., 2006; Duff and Lacks, 2008). After this size dependent charging, the smallest particles are separated from the larger ones by gravitation (smallest particles being bowled easily than the larges ones) inducing what might be called gravitational charge separation, which is consistent with the PG observations in the dust storms source region (e.g., Williams et al., 2009; Kamra, 1972). Nevertheless, the contact and triboelectric charging depends strongly on the grain collision frequency and though high frequencies are expected in dust storms near to the source region to cause dust electrification, this is not the case for regions far from it, as is the present case. The layers that reach distant locations have low densities; which corresponds to low collision frequencies and ruling out contact and triboelectric charging as charging mechanism. Assuming that dust charge decays in time-scales of minutes (Nicoll et al., 2011) means that the dust layers will loss its negative charge if charging is absent. Consequently, the dust layers will no longer be charged when away from the source regions. Thus, to explain long-range electrification it is reasonably to consider that the charging of the dust layers is caused by the action of the air-Earth electric current. In accordance to the discussion in Nicoll et al. (2011), a layer of uncharged dust particles will scavenge atmospheric ions by attachment to the large dust particles, reducing the air conductivity in that region. Such reduction in conductivity results in the creation of a space charge density (SCD) by the action of the air-Earth current as follows: air-Earth electric current, flowing from the Ionosphere to the Earth's surface, will bring to the upper part of the dust layer positive small ions that, after equilibrium is reached, will no longer be scavenged by the dust particles but will accumulate due to reduced air conductivity. This accumulation of charge will create a net positive SCD commonly represented by $\rho$ with a given dependence on the altitude, $z$, defining its vertical profile. If it is assumed that this vertical profile is defined in terms of the Heaviside function $H(z)$:

$$\rho(z) = \rho_0 [H(z-h) - H(z-h-t)]. \tag{1}$$

The effect on the PG is calculated straightforwardly by integration of Gauss' law. In Equation (1) it has been used $h$ as the height where the dust layer is located, $t$ is the thickness of the charge layer and $\rho_0$ is the space charge density. Previous observations of desert dust in Southern Portugal set $h \sim 3$ km (Obregón et al., 2015). Denoting the positive PG near the surface as $F$, Gauss' law relates the vertical variation of $F$ with $\rho(z)$ through the relation:

$$\frac{dF}{dz} = \frac{-\rho(z)}{\varepsilon_0}, \tag{2}$$

here $\varepsilon_0$ is the permittivity of vacuum. This equation can be integrated easily to estimate the effect of the space charge density on $F$:

$$\int_{F_h}^{F_{h+t}} dF = \frac{-\rho_0}{\varepsilon_0} \int_h^{h+t} dz \ . \tag{3}$$

Taking into account that the measurements are performed in the ground, the increase of the PG due to the space charge is defined by $\Delta F = F_h - F_{h+t}$, resulting in:

$$\Delta F = \frac{\rho_0}{\varepsilon_0} t . \tag{4}$$

Using Equation (4) and the observed $\Delta F \sim 23$ V/m (previous section) and charge layer thicknesses ranging $t \sim 10\text{-}100$ m, the values for the space charge density are estimated to be found in $\rho_0 \sim 20\text{-}2$ pCm$^{-3}$. This is in reasonable agreement with the values observed experimentally (Nicoll et al., 2011). Nevertheless, it should be mentioned that this is a simplified model and for that reason it has several limitations. Two fundamental simplifications have been made, one is the assumption of a uniform space charge distribution and the other is that this space charge occupies a semi-infinite plane in the $x$ and $y$ coordinates. These simplifications tend to overestimate the real effect of the desert dust. Future work may consider a more realistic model assuming a disc of space charge having a distribution different from the uniform.

## 6. Conclusions

Long-range electrification of a desert dust event, traveling at an altitude of $\sim 3$ km, is for the first time observed with surface atmospheric electrical Potential Gradient measurements. A triangular array of field-mills, covering an area of $\sim 1000$ km$^2$, has been used aiming to remove the local perturbations imposed on the Potential Gradient, mainly by local sources of pollution or space charges. The observations point to a long-range electrification that might be caused by the action of the air-Earth current on the low conductivity region occupied by the dust layer, forming a space charge layer inside the dust layer. Based on that simple mechanism a formulation was derived. Considering a charge layer thicknesses in the range of $\sim 10\text{-}100$ m, the values for the space charge density are estimated to be $\rho_0 \sim 20\text{-}2$ pCm$^{-3}$. These values are comparable with values found on the literature. A final remark is made to the fact that the dust storm electrical signal is a weak one as a consequence of the weaknesses of the dust storm itself. The present manuscript can be a motivation to perform longer campaigns that may be able to capture stronger dust storms and permit a better correlation between the dust storm induced electrical perturbations measured at all the sensors of the array.

**Acknowledgements**

Experiments were accomplished during the field campaign funded by FCT (Portuguese Science and Technology Foundation) and FEDER-COMPETE: ALEX 2014 (EXPL/GEO-MET/1422/2013) FCOMP-01-0124-FEDER-041840. This work is also co-funded by the European Union through the European Regional Development Fund, framed in COMPETE 2020 (Operational Programme Competitiveness and Internationalisation) through the ICT project (UID/GEO/04683/2013) with reference POCI-01-0145-FEDER-007690. HGS and SNP are grateful for the support by FCT through the post-doc grants: SFRH/BPD/63880/2009 and SFRH/BPD/81132/2011, respectively. **The authors are grateful to John Chubb for his useful and insightful discussions over this topic. He shall be missed.** Maintenance work of the field-mill installed at Beja Air-base done by the Meteorology Department (Carlos Policarpo and his team) is here truly recognized.

**References**

Calle, C.I., Buhler, C.R., McFall, J.L., and Snyder, S.J. (2009). Particle removal by electrostatic and dielectrophoretic forces for dust control during lunar exploration missions, Journal of Electrostatics 67, 89-92.

Chalmers, J. A. (1967), Atmospheric Electricity, 2nd ed., Pergamon, Oxford, U. K.

Chubb, J. (2015) Comparison of atmospheric electric field measurements by a pole mounted fieldmeter and by a horizontal wire antenna. Journal of Electrostatics 73, 1–5. DOI: 10.1016/j.elstat.2014.10.003.

Chubb, J. (2014). The measurement of atmospheric electric fields using pole mounted electrostatic fieldmeters. Journal of Electrostatics 72, 295–300. DOI: 10.1016/j.elstat.2014.05.002

Conceição, R. and Silva, H.G. (2015). Simulations of the Global Electrical Circuit coupled to local Potential Gradient measurements, Journal of Physics: Conference Series 646, 012017.

Conceição, R., Melgão, M., Silva, H.G., Nicoll, K., Harrison, R.G., and Reis, A.H. (2015). Transport of the smoke plume from Chiado's fire in Lisbon (Portugal) sensed by atmospheric electric field measurements. Air Quality, Atmosphere & Health 8. DOI: 10.1007/s11869-015-0337-4.

Delory, G.T., Farrell, W.M., Atreya, S.K., Renno, N.O., Wong, A.-S., Cummer, S.A., Sentman, D.D., Marshall, J.R., Rafkin, S.C.R., Catling, D.C. (2006). Oxidant Enhancement in Martian Dust Devils and Storms: Storm Electric Fields and Electron Dissociative Attachment. Astrobiology 6(3), 451-462.

Duff, N. and Lacks, D.J., (2008). Particle dynamics simulations of triboelectric charging in granular insulator systems. J. Electrost. 66, 51.

Engelstaedter, S., Tegen, I., and Washington, R. (2006). North African dust emissions and transport. Earth-Sci. Rev. 79 73–100.

Esposito, F.; Debei, S.; Bettanini, C.; Molfese, C.; Arruego Rodriguez, I.; Colombatti, G.; Harri, A. M.; Montmessin, F.; Wilson, C.; Aboudan, A.; Abbaki, S.; Apestigue, V.; Bellucci, G.; Berthelier, J. J.; Brucato, J. R.; Calcutt, S. B.; Cortecchia, F.; Cucciarrè, F.; Di Achille, G.; Ferri, F.; Forget, F.; Friso, E.; Genzer, M.; Haukka, H.; Jimènez, J. J.; Jimènez, S.; Josset, J. L.; Karatekin, O.; Landis, G.; Lorenz, R.; Marchetti, E.; Martinez, J.; Marty, L.; Mennella, V.; Möhlmann, D.; Moirin, D.; Molinaro, R.; Palomba, E.; Patel, M.; Pommereau, J. P.; Popa, C. I.; Rafkin, S.; Rannau, P.; Renno, N. O.; Schipani, P.; Schmidt, W.; Segato, E.; Silvestro, S.; Simoes, F.; Spiga, A.; Valero, F.; Vázquez, L.; Vivat, F.; Witasse, O.; Mugnuolo, R.; Pirrotta, S., The DREAMS Experiment of the ExoMars 2016 Mission for the Study of Martian Environment During the Dust Storm Season, Eighth International Conference on Mars, held July 14-18, 2014 in Pasadena, California. LPI Contribution No. 1791, p.1246 (2014).

Ette, A.I. (1970). The effect of the Harmattan dust on atmospheric electric parameters. J. Atmos. Terr. Phys. 33 295–300.

Freier, G.D. (1960). The electric field of a large dust devil, J. Geophys. Res., 65(10), 3504.

Grinsted, A., J. C. Moore, S. Jevrejeva (2004), Application of the cross wavelet transform and wavelet coherence to geophysical time series, Nonlin. Process. Geophys., 11, 561566.

Harrison, R. G. (2013). The Carnegie Curve. Surveys in Geophysics, 34, 209-232.

Holben, B.N., Eck, T.F., Slutsker, I. et al. (1998). AERONET— A federated instrument network and data archive for aerosol characterization. Remote Sensing of Environment, 66 (1), 1–16. DOI: 10.1016/S0034-4257(98)00031-5

Inculet, I.I., Castle, G.S.P., Aartsen, G. (2006). Generation of bipolar electric fields during industrial handling of powders. Chem. Eng. Sci. 61, 2249–2253.

Kamra, A. (1972). Measurements of the electrical properties of dust storms. J. Geophys. Res. 77 5856–69.

Lopes, F., Silva, H.G., Bárias, S., and Barbosa, S.M. (2015) Preliminary results on soil-emitted gamma radiation and its relation with the local atmospheric electric field at Amieira (Portugal). Journal of Physics: Conference Series 646, 012015. DOI: 10.1088/1742-6596/646/1/012015.

Lopes, F., Silva, H.G., Salgado, R., Potes, M., Nicoll, K., and Harrison, G. (2016). Atmospheric Electrical Field measurements near a Fresh Water Reservoir and the formation of the Lake Breeze (Tellus a, submitted).

Nicoll, K.A., and Harrison, R.G. (2010). Experimental determination of layer cloud edge charging from cosmic ray ionisation. Geophys. Res. Lett. 37, L13802.

Nicoll, K.A., Harrison, R.G., and Ulanowski, Z. (2011). Observations of Saharan dust layer electrification, Environ. Res. Lett. 6, 014001.

Obregón, M.A., Pereira, S., Salgueiro, V., Costa, M.J., Silva, A.M., Serrano, A., and Bortoli, D. (2015). Aerosol radiative effects during two desert dust events in August 2012 over the Southwestern Iberian Peninsula. Atmospheric Research 153, 404–415. DOI: 10.1016/j.atmosres.2014.10.007

Preißler, J., F. Wagner, J. L. Guerrero-Rascado, and A. M. Silva (2013). Two years of free-tropospheric aerosol layers observed over Portugal by lidar, J. Geophys. Res. Atmos., 118, 3676–3686.

Reiter, R. (1992). Phenomena in Atmospheric and Environmental Electricity. Developments in Atmospheric Sciences, Elsevier (562 pp.).

Rycroft, M. J., Israelsson, S. and Price, C. (2000). The global atmospheric electric circuit, solar activity and climate change Jour. Atmosph. Solar-Ter. Phys. 62 1563-1576.

Rudge, W.A.D. (1913). Atmospheric Electrification during South African Dust Storms. Nature 91, 31-32.

Sarver, T., Qaraghuli, A.A., and Kazmerski, L.L. (2013). A comprehensive review of the impact of dust on the use of solar energy: History, investigations, results, literature, and mitigation approaches. Renewable and Sustainable Energy Reviews 22, 698.

Silva, H.G., Conceição, R., Melgão, M., Nicoll, K., Mendes, P.B., Tlemçani, M., Reis, A.H., and Harrison, R.G. (2014). Atmospheric electric field measurements in urban environment and the pollutant aerosol weekly dependence, Environment Research Letters, 9, 114025.

Siingh, D., Singh, R.P., Gopalakrishnan, V., Selvaraj, C., and Panneerselvam, C. (2013). Fair-weather atmospheric electricity study at Maitri (Antarctica). Earth Planets Space 65, 1541–1553.

Stein, A.F., Draxler, R.R., Rolph, G.D., Stunder, B.J.B., Cohen, M.D., and Ngan, F. (2015). NOAA's Hysplit atmospheric transport and dispersion modeling system. Bull. Amer. Meteor. Soc., 96, 2059–2077.

Tacza, J., Raulin, J.-P., Macotela, E., Norabuena, E., Fernandez, G., Correia, E., Rycroft, M.J., and Harrison, R.G. (2014). A new South American network to study the atmospheric electric field and its variations related to geophysical phenomena. J. Atmos. Sol-Terr. Phys., 120, 70–79. DOI: 10.1016/j.jastp.2014.09.001.

Torrence, C., and Compo, G.P. (1998). A practical Guide to Wavelet Analysis. Bull. Amer. Meteor. Soc. 79(1), 61-78.

Ulanowski, Z., Bailey, J., Lucas, P.W., Houghm J.H., Hirst, E. (2007) Alignment of atmospheric mineral dust due to electric field. Atmos. Chem. Phys. 7 6161–73.

Voeikov, A.I. (1965). IInstruction on Preparation of the Material and Publication of the results of Atmospheric Electric ObservationsQ, Ed. Main Geophysical Observatory, Leningrad.

Williams, E., Nathou, N., Hicks, E., Pontikis, C., Russell, B., Miller, M., and Bartholomew, M.J. (2009). The electrification of dust-lofting gust fronts ('haboobs') in the Sahel, Atmos. Res., 91, 292-298.

Winker, D. M., W. H. Hunt, and M. J. McGill (2007), Initial performance assessment of CALIOP, Geophys. Res. Lett., 34, L19803, doi:10.1029/2007GL030135.

**Figure 1. A)** Global reference with the position for the capitals of Portugal (Lisbon), Spain (Madrid) and Morocco (Rabat). **B)** Geographic location of the three sites used for measurements of atmospheric electric potential gradient: Évora (EVO, 38.50º N, 7.91º W), Amieira (AMI, 38.27º N, 7.53º W) and Beja (BEA, 38.07º N, 7.93º W). The three measuring sites are separated by ~ 50 km, forming approximately an equilateral triangle with an area of ~ 1000 km². The small yellow rectangle represents an area in the South of Spain where Granada, Malaga and Cerro de Poyos AERONET stations are located.

**Figure 2.** (a) Aerosol optical depth (at 440 nm); (b) Angstrom exponent for Évora and Badajoz, as well as for three stations in the south of Spain, namely Garanada, Malaga and Cerro de Poyos.

**Figure 3.** NAAPS maps of total optical depth, from 16 July 2014 at 00:00 UTC (a) to 18 July 2014 12:00 UTC (f), every 12 hours.

**Figure 4.** Attenuated backscatter coefficient (a), depolarization ratio (b), vertical feature mask (c) and aerosol classification (d) data from CALIOP (Cloud Aerosol Lidar with Orthogonal Polarization) aboard of CALIPSO (Cloud-Aerosol Lidar and Infrared Pathfinder Satellite Observations).

**Figure 5.** Atmospheric Electric Potential Gradient in Évora (EVO): (a) raw data (back solid line represents a robust lowess smoothing); (b) wavelet periodogram (1-hour averaged). The vertical dashed line in both panels represents the 17th of July 2014 desert dust event. The gaps correspond to missing data.

**Figure 6.** Atmospheric Electric Potential Gradient in Amieira (AMI): (a) raw data (back solid line represents a robust lowess smoothing); (b) wavelet periodogram (1-hour averaged). The vertical dashed line in both panels represents the 17th of July 2014 desert dust event. The gaps correspond to missing data.

**Figure 7.** Atmospheric Electric Potential Gradient in Beja (BEA): (a) raw data (back solid line represents a robust lowess smoothing); (b) wavelet periodogram (1-hour averaged). The vertical dashed line in both panels represents the 17$^{th}$ of July 2014 desert dust event. The gaps correspond to missing data.

**Figure 8.** Upper, middle and lower five panels represent the daily variation for EVO, AMI and BEA stations, respectively, for days 44 to 48 of the ALEX2014 campaign. The thick black lines in the plots represent the lowess smoothed mean daily variation of the PG for the all campaign in the respective stations. The Saharan dust event starts on day 45 (15th July 2014) and ends on day 47 (17th July 2014).

**Figure 9.** Daily boxplot representation of: (a) Angstrom Exponent; (b) Potential Gradient on EVO; (c) Potential Gradient on AMI; (d) Potential Gradient on BEA. The vertical dashed line marks the beginning of the desert dust event and the solid curves on the PG data represent the lowess smoothing of that data.

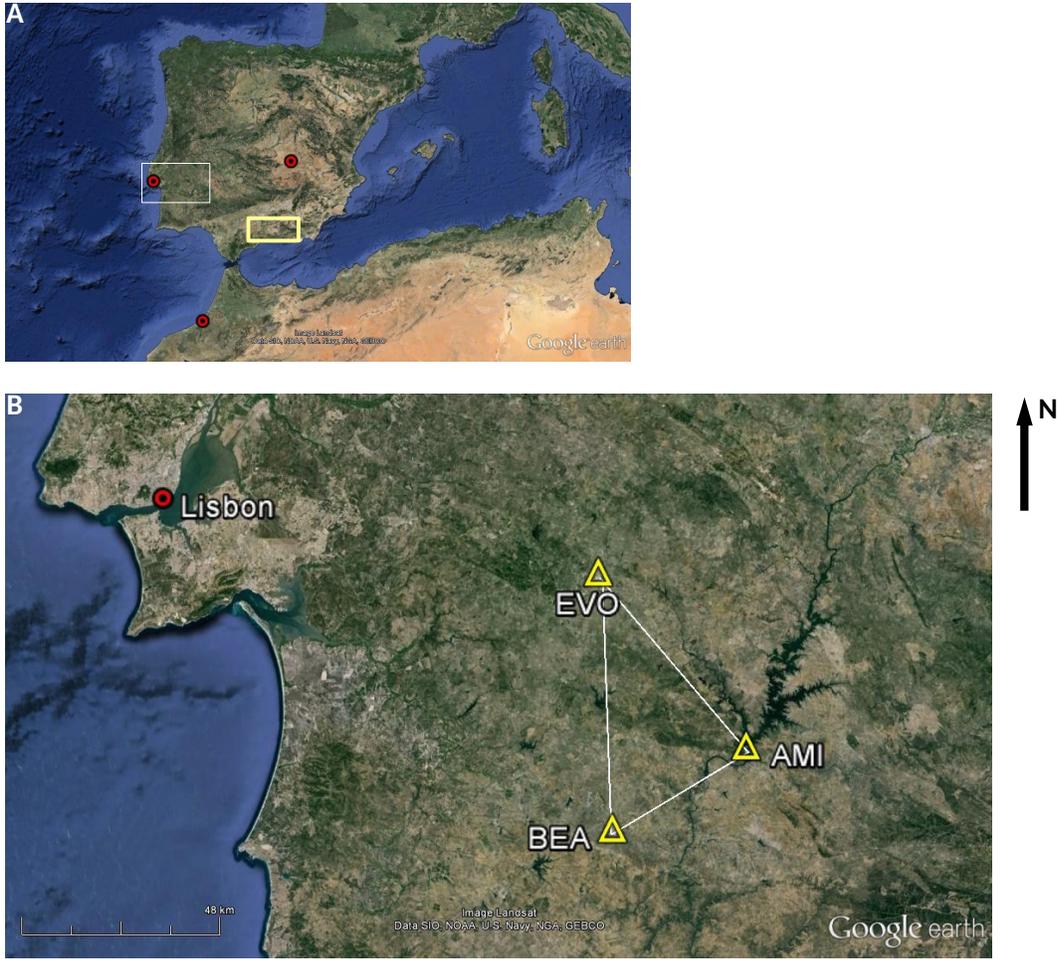

**Figure 1**

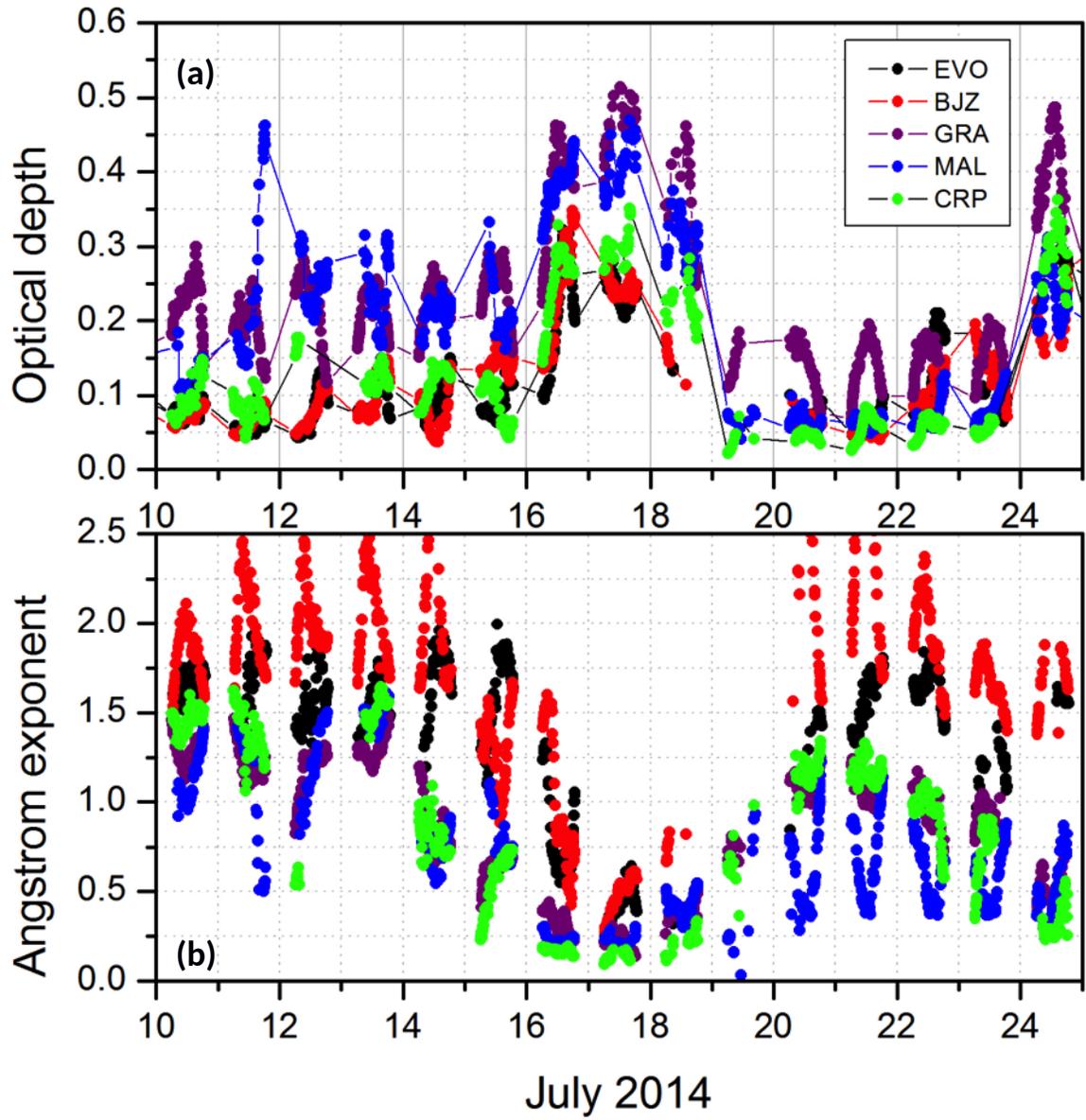

**Figure 2**

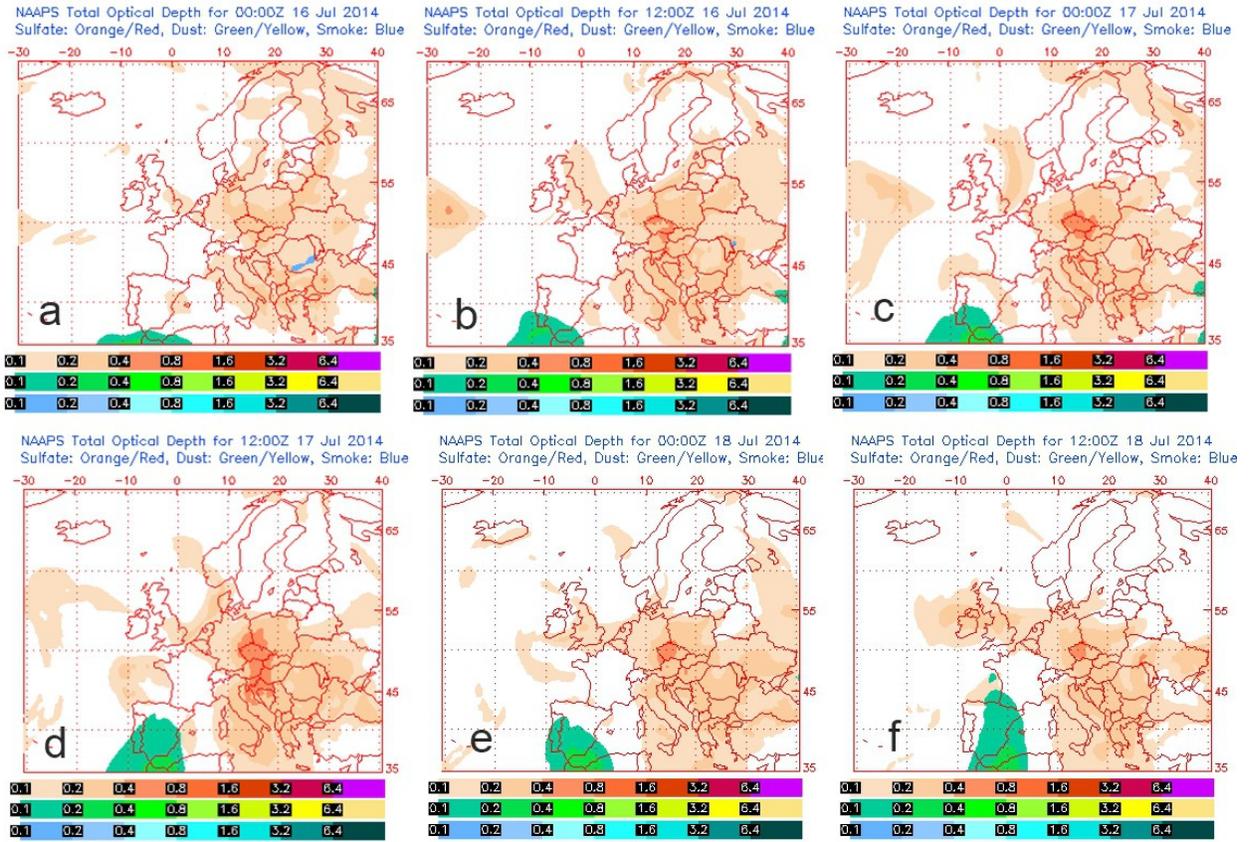

**Figure 3**

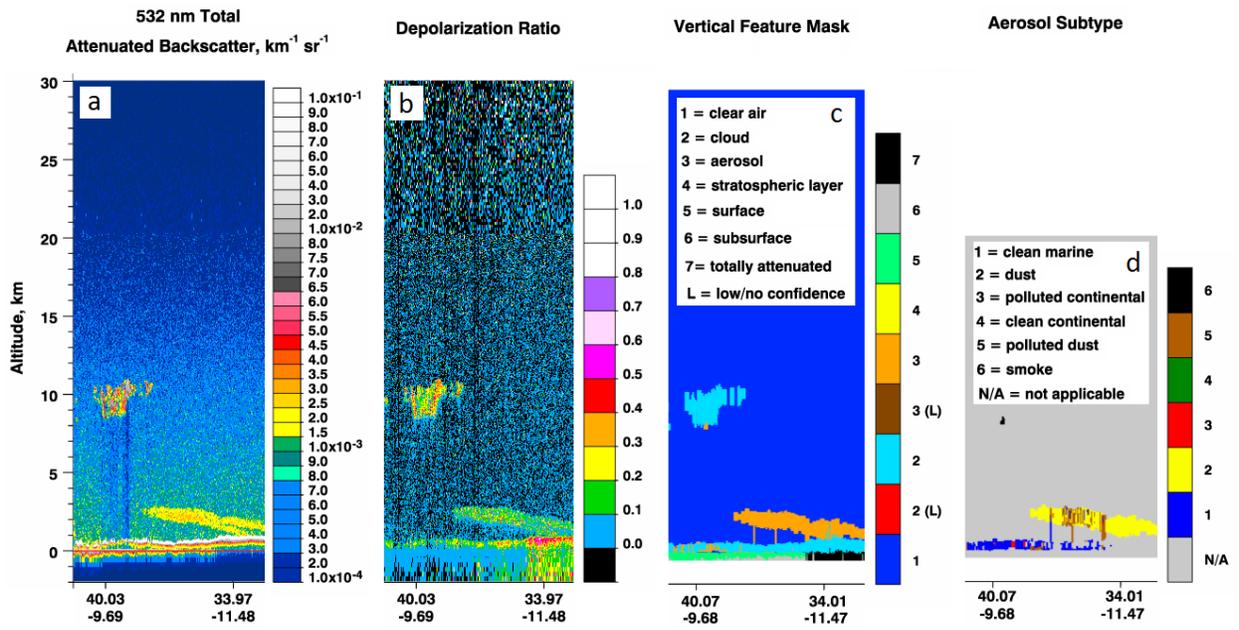

**Figure 4**

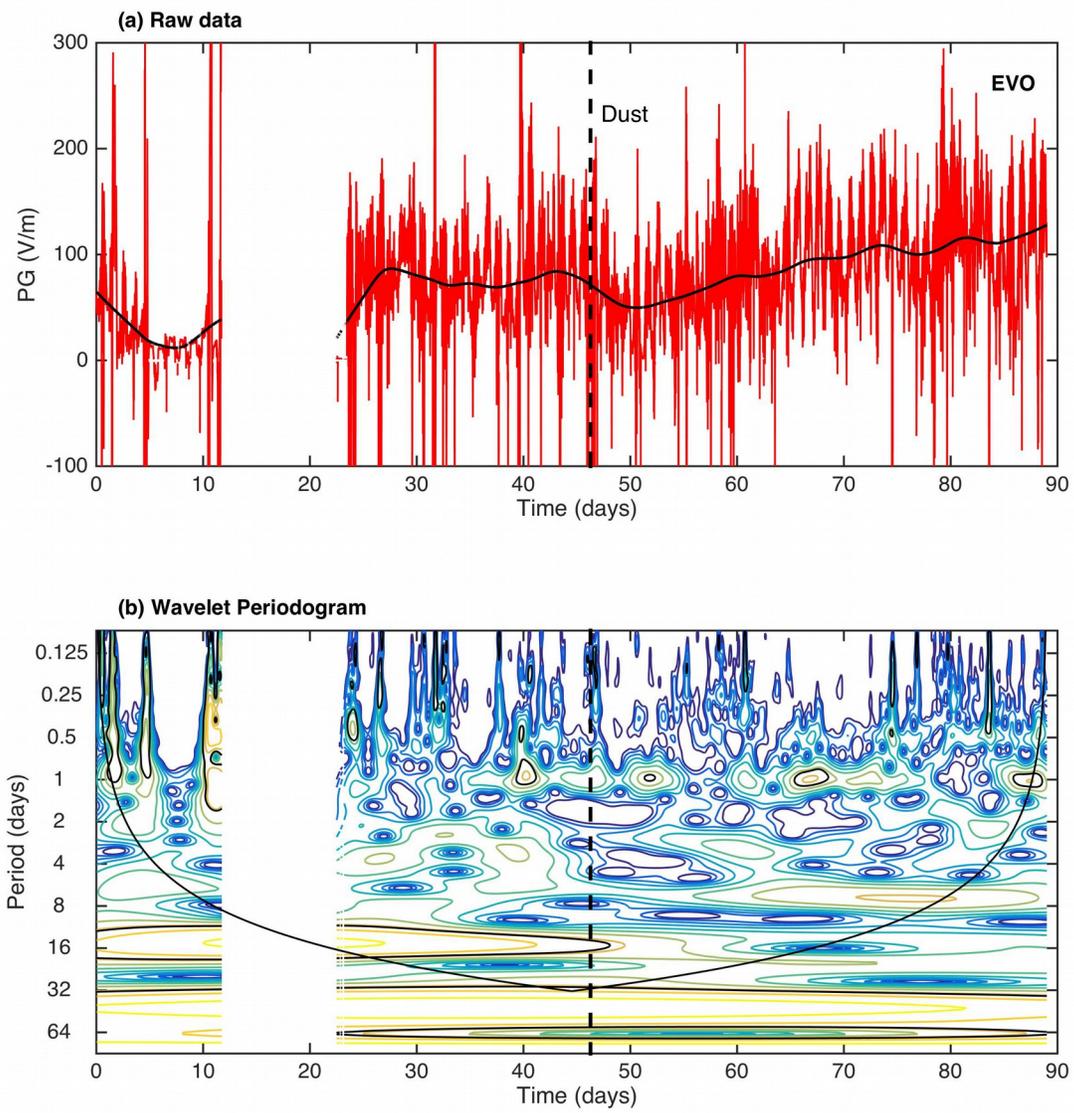

**Figure 5**

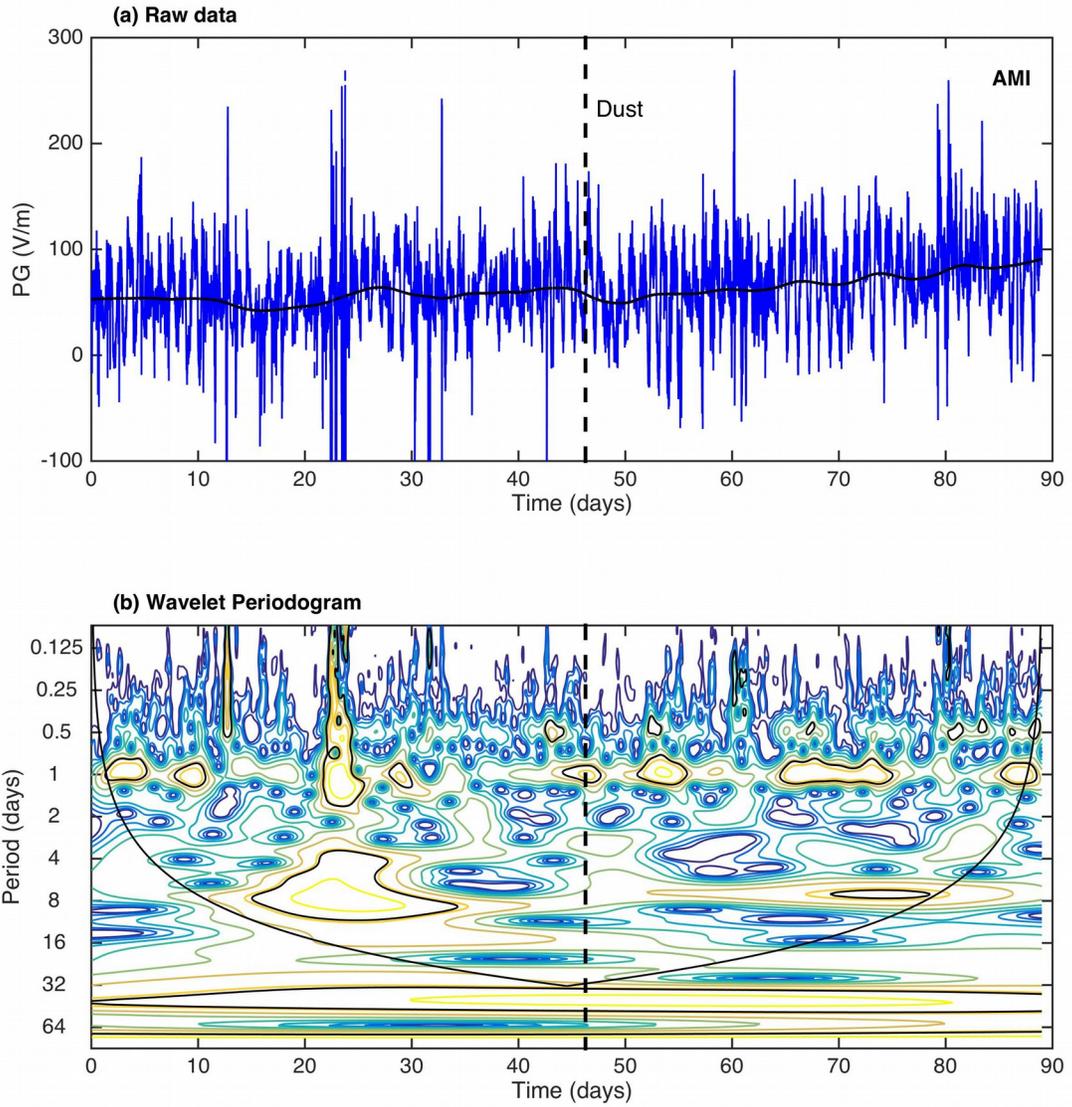

**Figure 6**

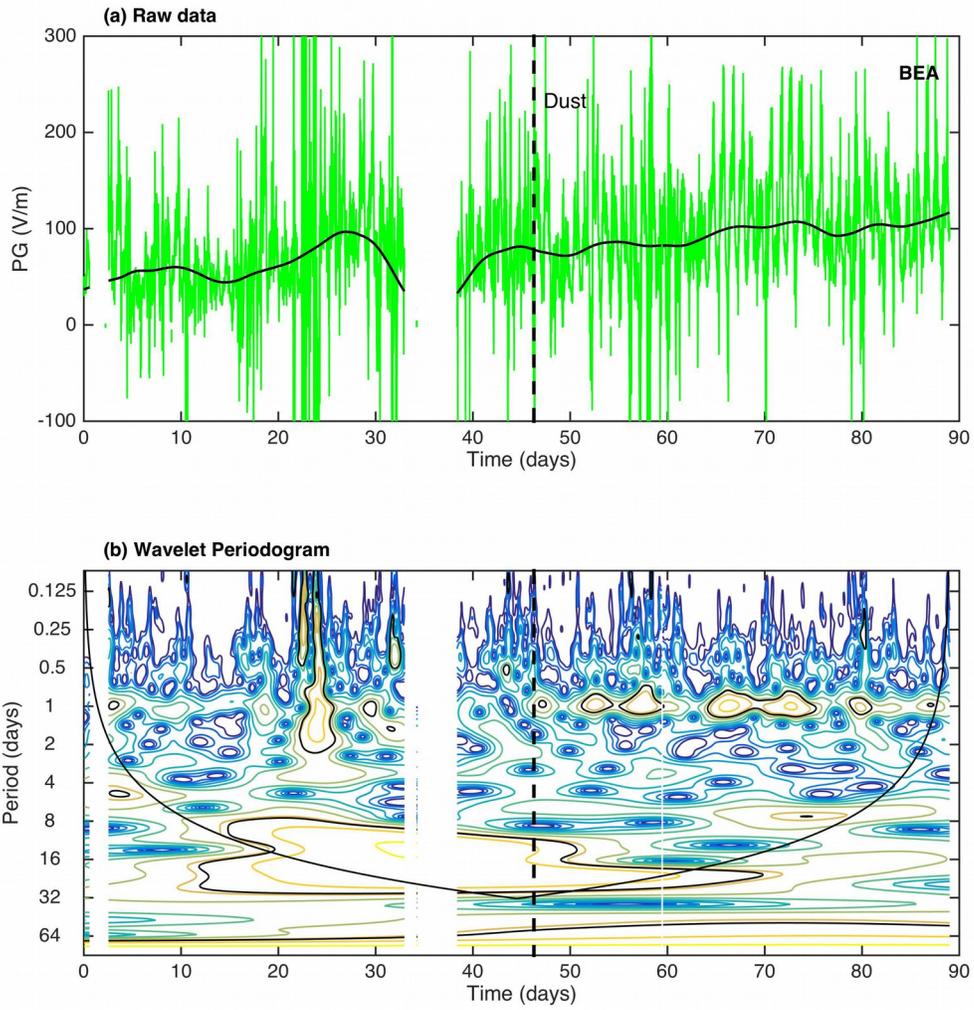

**Figure 7**

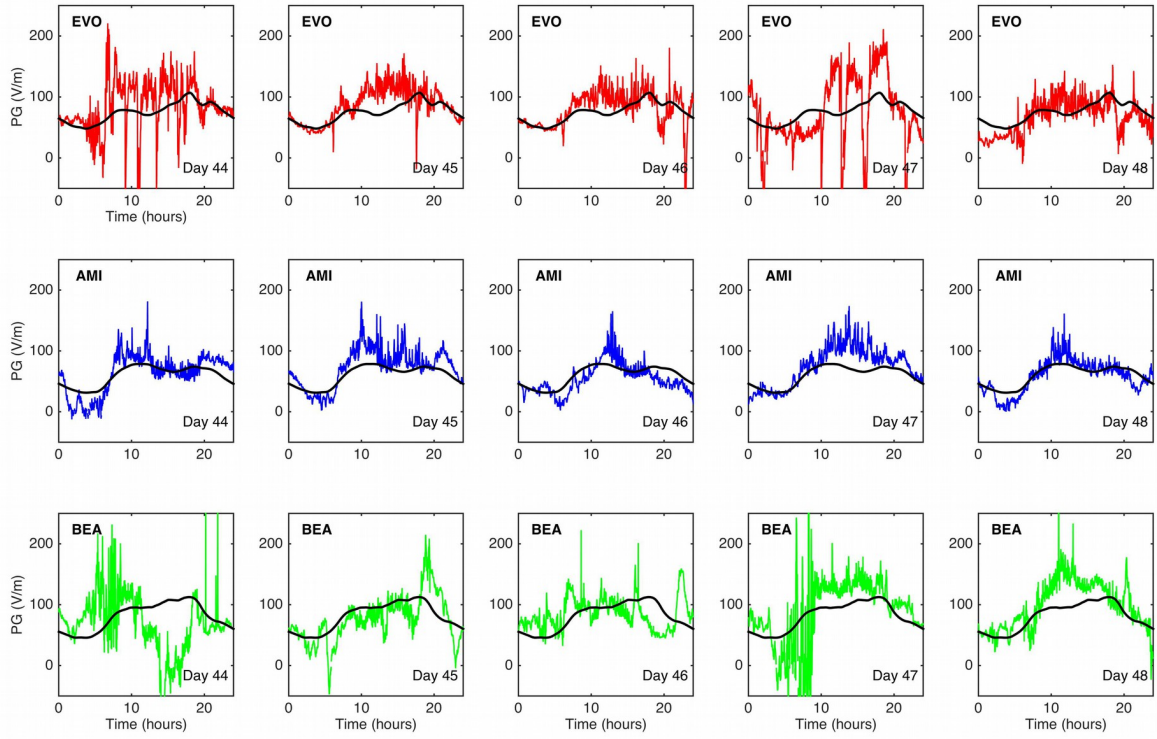

**Figure 8**

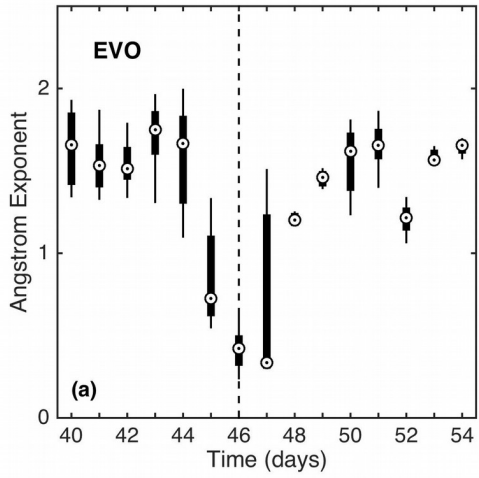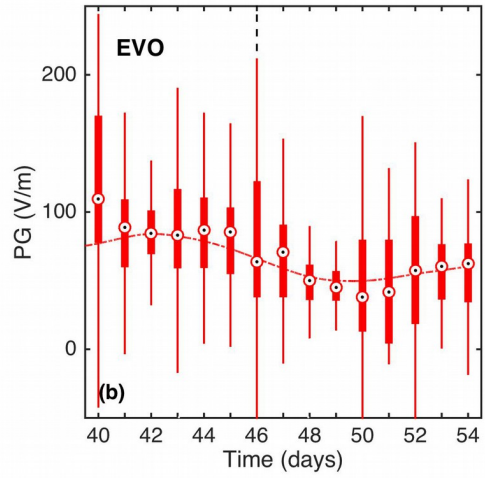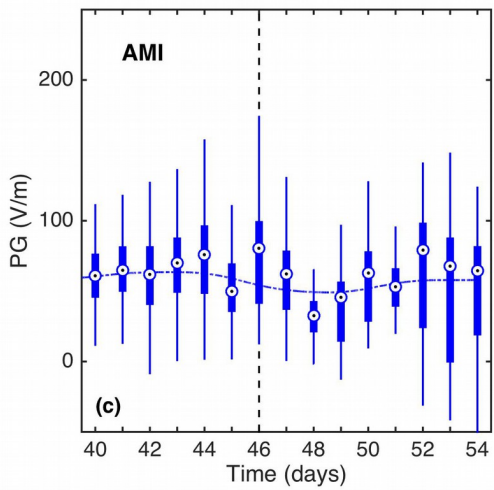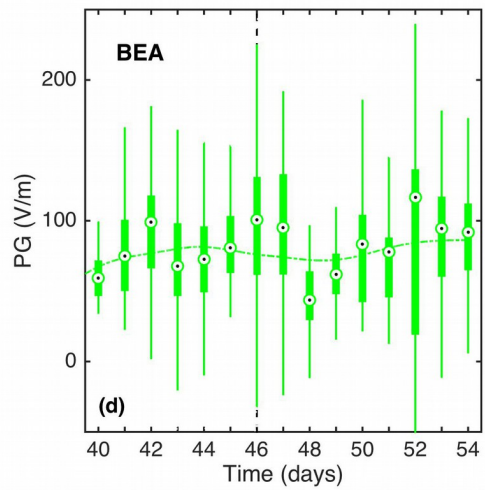

**Figure 9**